\documentclass[pre,superscriptaddress,twocolumn,nofootinbib]{revtex4-2}

\usepackage{graphicx}
\usepackage{color}
\usepackage[usenames,dvipsnames]{xcolor}
\usepackage{amsmath,bbm,amssymb, amsthm}
\usepackage{dsfont} 
\usepackage{stmaryrd}
\usepackage{comment}
\usepackage{numprint}

\def\bb{\begin{eqnarray}}
\def\ee{\end{eqnarray}}
\newcommand{\ket}[1]{| #1 \rangle}
\newcommand{\bra}[1]{\langle #1 |}

\newcommand{\moy}[1]{\left\langle #1 \right\rangle} 
\newcommand{\idop}{\mathds{1}}

\begin{document}

\title{Revealing the fuel of a quantum continuous measurement-based refrigerator}

\author{Cyril Elouard}
\email{cyril.elouard@univ-lorraine.fr} 
\affiliation{Université de Lorraine, CNRS, LPCT, F-54000 Nancy, France}

\author{Sreenath K. Manikandan}
\altaffiliation[Current affiliation: ]{Tata Institute of Fundamental Research Hyderabad, 36/P, Gopanpally Village, Serilingampally Mandal, Hyderabad, Telangana 500046, India}
	\affiliation{Nordita,
KTH Royal Institute of Technology and Stockholm University,
Hannes Alfv\'{e}ns v\"{a}g 12, SE-106 91 Stockholm, Sweden}

\author{Andrew N. Jordan}
\affiliation{Institute for Quantum Studies, Chapman University, Orange, California, USA}
\affiliation{The Kennedy Chair in Physics, Chapman University, Orange, CA 92866, USA}
\affiliation{Department of Physics and Astronomy, University of Rochester, Rochester, NY 14627, USA}
\author{G\'eraldine Haack}
\email{geraldine.haack@unige.ch} 
\affiliation{Department of Applied Physics, University of Geneva, 1211 Genève, Switzerland}

\begin{abstract}
   While quantum measurements have been shown to constitute a resource for operating quantum thermal machines, the nature of the energy exchanges involved in the interaction between system and measuring apparatus is still under debate. In this work, we show that a microscopic model of the apparatus is necessary to unambiguously determine whether quantum measurements provide energy in the form of heat or work. We illustrate this result by considering a measurement-based refrigerator, made of a double quantum dot embedded in a two-terminal device, with the charge of one of the dots being continuously monitored. Tuning the parameters of the measurement device interpolates between a heat- and a work-fueled regimes with very different thermodynamic efficiency. Notably, we demonstrate a trade-off between a maximal thermodynamic efficiency when the measurement-based refrigerator is fueled by heat and a maximal measurement efficiency quantified by the signal-to-noise ratio in the work-fueled regime. Our analysis offers a new perspective on the nature of the energy exchanges occurring during a quantum measurement, paving the way for energy optimization in quantum protocols and quantum machines.
\end{abstract}

\maketitle

\section{Introduction} 
In the quantum world, quantum measurements change the state of the measured system. This measurement-induced dynamics can exhibit entropy and energy transfers between the measuring apparatus and the system \cite{Elouard17Role}, which in turn are witnesses of the nontrivial thermodynamic balance behind the measurement process. On the one hand, the minimum work cost of measurement has been shown to strongly depend on measurement performance \cite{Sagawa09,Latune25}, leading to a diverging resource cost for ideal (i.e. projective) measurements \cite{Guryanova20}. On the other hand, while a measurement can be shown to be a source of work under ideal conditions \cite{Jacobs09,Elouard17}, the stochasticity of the measurement-induced dynamics in the general case is reminiscent of the action of a thermal bath \cite{Elouard17Role}, and can even induce thermalization \cite{Ashida18}. Moreover, several theories attempting to describe the emergence of the non-unitary measurement-induced dynamics have emphasized the key role of the inaccessible degrees of freedom of the measuring apparatus \cite{Zurek09,Elouard21}, which in thermodynamic language constitute a heat bath \cite{Allahverdyan13,Latune25}, and are responsible for irreversibility and heat transfers occurring during the measurement.

\begin{figure}
\includegraphics[width=0.45\textwidth]{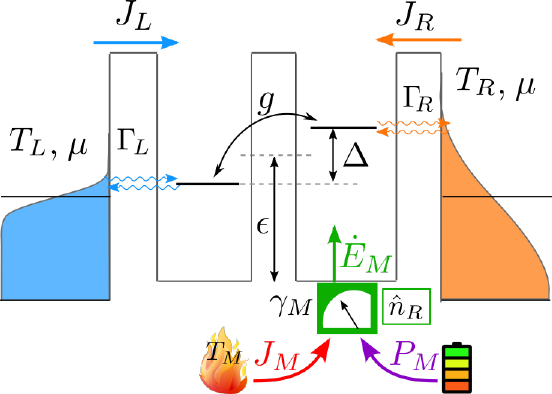}
\caption{Principle of the continuous measurement-powered refrigerator. The charge in the right quantum dot is continuously monitored by an apparatus at rate $\gamma_M$. As this measurement does not commute with the dot Hamiltonians, it results in an energy flow $\dot{E}_M$, which in turn powers heat transfer from the cold (left) to the hot (right) lead, that is $J_L < 0 < J_S$. On general grounds, the measuring apparatus can be powered either by a source of work providing power $P_M$ and/or a source of heat at temperature $T_M$ providing heat flow $J_M$.}\label{f:Setup}
\end{figure}

Elucidating the thermodynamic nature of the energy provided by a quantum measurement is a prerequisite to quantifying the efficiency of engine cycles incorporating quantum measurements such as quantum Maxwell demons \cite{Cottet17,Naghiloo18,Annby-Andersson24}, and measurement-powered machines \cite{Elouard17,Elouard18,Ding18,Buffoni19,Jordan20,Manikandan22,Jussiau23,Bresque21}. It is also a necessary step towards systematic energetic optimization of protocols involving a large number of quantum measurements such as fault-tolerant quantum algorithms \cite{Shor96}.
Here, we address this question and show that the nature of the energy transfer occurring during a quantum measurement is controlled by microscopic parameters of the measuring apparatus, and can be tuned from a regime where the measurement is mostly fueled by heat, to a regime where it is mostly fueled by work. When the measurement is incorporated into an engine cycle, these regimes correspond to very different values of the thermodynamic efficiency for the machine, even if the average measurement-induced dynamics appears to be the same.

We illustrate this on an example of measurement-powered refrigerator based on two single-electron quantum-dots (QDs) coupled to electron reservoirs at different temperatures. Unlike a previous proposal of a two-qubit measurement-based refrigerator \cite{Buffoni19}, our machine involves a measurement that is local (rather then occurring in an entangled basis) and is in competition with an always-on tunnel coupling between the QDs. Moreover, instead of using a discrete projective measurement as a stroke of our machine cycle (which would correspond to infinite work cost \cite{Guryanova20}), we consider a weak continuous measurement \cite{Jordan24} which results in a steady state cooling power \cite{Bhandari22}. A similar model was proposed as an engine in \cite{Bettmann23}. 

Conducting a thermodynamic analysis of the device, we show that its efficiency depends on the nature of its fuel, heat, work or a mixture of both. It is only by considering a microscopic model of the measuring apparatus that we can distinguish between work-fueled and heat-fueled regimes, and investigate without ambiguity the performances of this measurement-based refrigerator. As realistic charge measuring apparatus, we consider a biased electronic tunnel junction -- a quantum point contact (QPC) -- capacitively coupled to one of the QD. Remarkably, we highlight a trade-off between the efficiency of the machine, which is maximized in the heat-fueled regime, and the signal-to-noise ratio which becomes maximal when the QPC is powered by work input only. The work performed to operate the machine is then mostly dissipated into the reservoirs connected to the QPC, leading to poor cooling efficiency. Our results show that determining the nature of energy exchanges do require a microscopic knowledge of the device, and that work-fueled devices do not necessarily correspond to the best thermodynamic efficiency.\\

\section{A continuous measurement-fueled electronic refrigerator} 

\subsection{Model}

We consider as working body a system composed of two tunnel-coupled single energy-level QDs, each coupled to an electron reservoir. In addition, we assume that the number of electrons in the right-hand dot is continuously monitored. This setup is illustrated in Fig.~\ref{f:Setup}. We restrict the analysis to weak-system bath coupling, such that the dynamics and steady-state are well captured by a master equation (ME), and to the regime of strong inter-dot Coulomb repulsion, such that the dynamics is confined to the single-electron Hilbert subspace of the dots, described by the Hamiltonien
\bb
H_\text{QD} &=& (\epsilon+\Delta/2)c_L^\dagger c_L + (\epsilon-\Delta/2)c_R^\dagger c_R \nonumber\\
&&+ (g/2)(c_L^\dagger c_R + c_R^\dagger c_L),
\ee
with $c_L = \ket{00}\bra{10}$ ($c_R=\ket{00}\bra{01}$) the operators annihilating an electron in the left (right) dot. $\ket{00}$, $\ket{10}$ and $\ket{01}$ denote the states where both dots are empty, the left dot is occupied, and the right dot is occupied, respectively. Moreover, $\epsilon$, $\Delta$ and $g$ are, respectively, the average charging energy, the detuning and the inter-dot tunnel coupling. Explicit calculations detailed in Appendix \ref{s:App_ME} show that a valid regime for operating this device as a measurement-fueled refrigerator corresponds to a strong inter-dot tunnel coupling $g$ with respect to the typical system-bath coupling $\gamma$, captured by a global ME. The latter involves transitions between the global energy eigenstates of the QDs system, i.e. the doubly empty state $\ket{00}$ (energy $0$), and the hybridized states 
\bb
\ket{+} &=& \cos(\theta)\ket{10} + \sin(\theta)\ket{01} \,,\nonumber\\
\ket{-} &=& -\sin(\theta)\ket{10} + \cos(\theta)\ket{01}\,.
\ee 
(energies $\epsilon \pm \Omega/2$, with $\Omega=\sqrt{\Delta^2+g^2}$), with the angle theta defined via $\tan\theta = \sqrt{\frac{\Omega-\delta}{\Omega+\delta}}$.

The ME for the reduced density operator $\rho$ of the QDs takes the form:
\bb
\label{eq:ME}
\dot\rho = -i[H_\text{QD},\rho] + \big( {\cal L}_L  + {\cal L}_R + \cal{L}_M \big) \rho, 
\ee
with three dissipation terms defined below. First,
\bb
{\cal L}_\alpha &=& \sum_{l = \pm} \,\left(\Gamma_{\alpha l\downarrow}{\cal D}[c_l] +  \Gamma_{\alpha l \uparrow}\mathcal{D}[c_l^\dagger]\right), \quad \alpha = L,R\,,\label{eq:LLR} 
\ee
with $c_\pm = \ket{00}\bra{\pm}$ and  with $\mathcal{D}[X]\bullet \equiv X\bullet X^\dagger - \tfrac{1}{2}(X^\dagger X\bullet + \bullet X^\dagger X)$, captures the dissipation to the left and right reservoirs.  The exchanges of particles with the reservoirs are set by strengths $\Gamma_{\alpha l \uparrow} =\gamma f_{\alpha l}$ and $\Gamma_{\alpha l,\downarrow} =\gamma \left(1-f_{\alpha l})\right)$, with $f_{\alpha l} = [ e^{(\epsilon+l\Omega/2-\mu_\alpha)/kT_\alpha}+1]^{-1}$ the Fermi function of lead $\alpha$ characterized by chemical potential $\mu_\alpha$ and temperature $T_\alpha$, and the bare rate $\gamma =  2\pi\sum_{k} g_{\alpha,k}^2\delta(\epsilon-\omega_k) \ll \epsilon$ that we assume equal for both reservoirs for the sake of simplicity. Moreover,
\bb
{\cal L}_M &=& \frac{1}{\Delta t}\int_t^{t+\Delta t} \!\!\!\!dt' \gamma_M \mathcal{D}[\hat n^I_R(t')] = \gamma_M \!\!\!\!\sum_{\omega=0, \pm \Omega} \!\!\!\! \mathcal{D}[\hat{n}_\omega]\,, \label{eq:LM}
\ee
phenomologically captures the continuous measurement of the right-dot charge $\hat n_R = \ket{01}\bra{01}$, with measurement rate $\gamma_M$ \cite{Jacobs06}. We stress that, as any Lindblad equation, the derivation of Eq.~\eqref{eq:ME} from the Hamiltonian microscopic model involved a coarse-graining in time with time step $\Delta t$ assuming $\Delta t \gg \Omega^{-1}$ \cite{CohenBook}, which also had to be taken into account when deriving the form of $\mathcal{L}_M$, see Eq.~\eqref{eq:LM}. This was done by assuming $\Delta t \gamma_M \ll 1$, and introducing the interaction-picture representation of the measured charge, $\hat{n}^I_R(t) = e^{iH_\text{QD}t}\hat n_R e^{-iH_\text{QD}t} = \sum_{\omega=0,\pm\Omega} \hat n_\omega e^{i\omega t}$. We refer again to the Appendix \ref{s:App_ME} for all technical details.\\

\subsection{Thermodynamic balance} 

We calculate steady-state energy and heat flows by solving the master equation Eq.~\eqref{eq:ME} at long times where $\dot{\rho}_{ss}=0$. We recall that steady-state heat flows provided by each reservoir $\alpha =L,R$ are defined as the difference between the energy flow {$\dot E_\alpha$} from this reservoir and the electrical power {$P_\alpha$} carried by exchanged particles due to a finite chemical potential $\mu_\alpha$ \cite{Benenti17}: 
\bb
J_\alpha &=& \dot{E}_\alpha - P_\alpha =\text{Tr}\{(H_\text{QD}- \mu_\alpha \sum_l c_l^\dagger c_l){\cal L}_\alpha \rho_{ss} \}\,.\;\;\;
\ee
In the following, we focus on the case $\mu_L=\mu_R=\mu$ (no net electric work exchanged with the reservoirs). We also introduce the flow of energy $\dot{E}_M$ provided by the measuring apparatus to the system, computed from:
\bb
\dot{E}_M = \text{Tr}\{H_\text{QD} {\cal L}_M \rho_{ss} \},
\ee

\begin{figure}
\includegraphics[width=0.95\columnwidth]{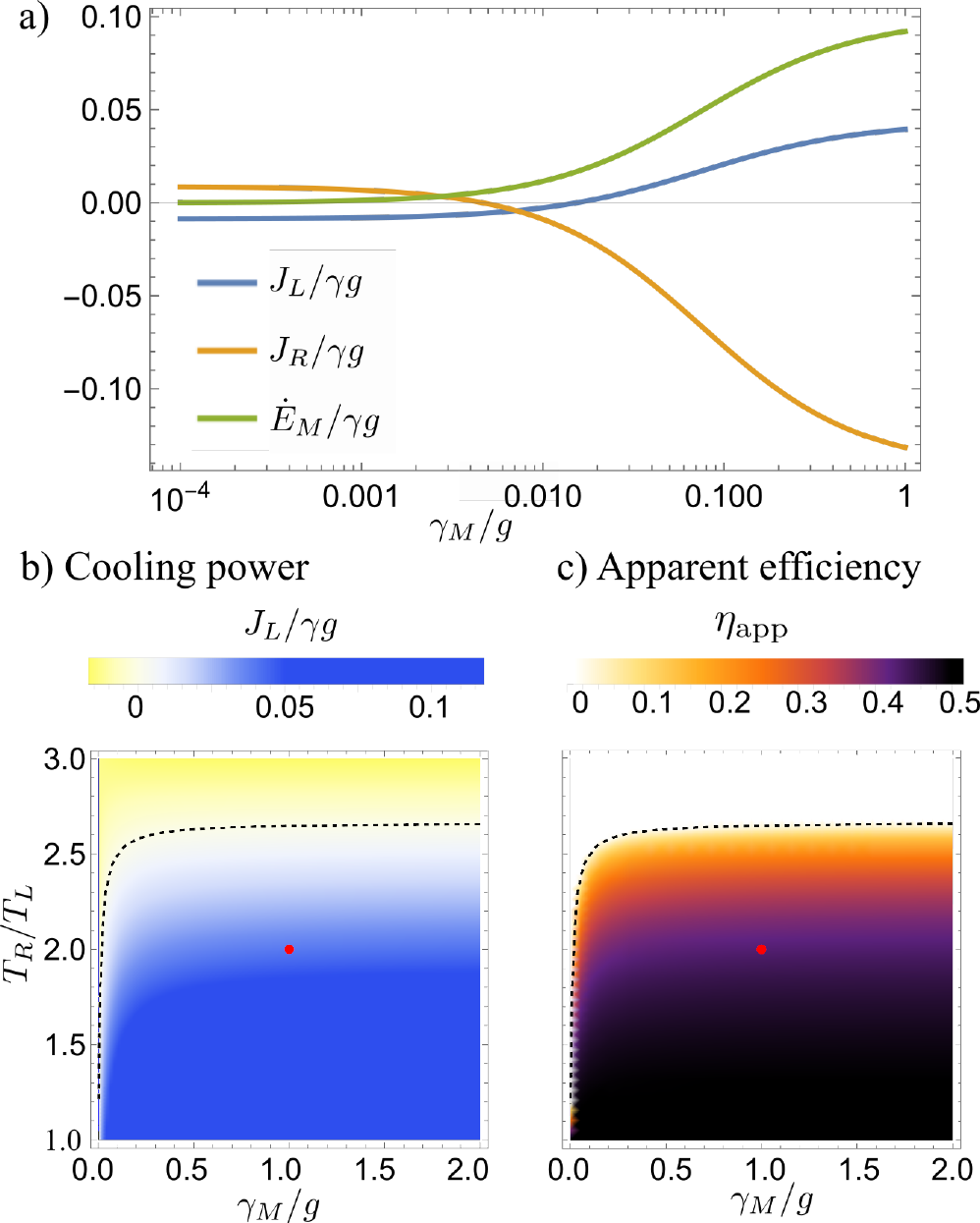}
\caption{a): Stationary cooling power $J_L$ (blue), heat flow received from the hot lead $J_R$ (orange) and energy flow provided by the measuring apparatus $\dot{E}_M$ (green) in unit of $\gamma g$ as a function of the measurement strength $\gamma_M$. b) Cooling power $J_L$ in units of $\gamma g$ as a function of the ratio of the hot over cold lead temperatures $T_R/T_L$ and the measurement strength $
\gamma/g$. c) Efficiency of conversion of measurement energy into cooling power $\eta_\text{id}=J_L/\dot{E}_M$  as a function of the ratio of the hot over cold lead temperatures $T_R/T_L$ and the measurement strength $
\gamma/g$. 
In a) and b) and c), other parameters are set as follows: $\gamma/g= 0.01$, $\Delta/g = 4.3$, $\mu/g = 10$, $\epsilon/g = 5.4$, $\gamma_M/g = 1$, $T_L/g = 2$, and for a): $T_R/T_L =2$. The red dots indicate the parameter values considered in the QPC model.}\label{f:currents}
\end{figure}

Energy conservation for the working body takes then the form:
\bb
\label{eq:first_law}
\frac{dU}{dt}= J_L + J_R + \dot{E}_M =0\,,
\ee
with $U = \text{Tr}\{H_\text{QD}\rho_{ss}\}$ the steady-state value of working body internal energy. \\

Assuming $T_R > T_L$ for the rest of this work, this device operates as a refrigerator whenever $J_L > 0$ (see Fig.~\ref{f:Setup} for the sign convention). In Fig.~\ref{f:currents}a), we show the heat flows $J_L, J_R$ and energy flow $\dot{E}_M$ as a function of the measurement strength $\gamma_M$. At $\gamma_M=0$, the measuring apparatus is not active, and the device reduces to a simple double dot subject to a thermal bias.  Consequently, the heat flows $J_L$ and $J_R$ have the same magnitude and opposite directions, defining a heat current from hot to cold, as expected. No cooling process takes place. At sufficiently large value of $\gamma_M/g$, and for a fixed temperature bias $\Delta T = T_R - T_L$ not exceeding a critical value $\Delta T_c >0$, energy provided by the measuring apparatus $\dot{E}_M >0$ (green curve) is used to continuously extract heat from the cold bath ($J_L>0$, blue curve) and to dissipate it into the hot bath ($J_R <0$, orange curve). The device then operates as a measurement-fueled refrigerator. Without any further prescription to analyze the nature of the energy flow from the apparatus in terms of heat or work, we can only compute an \textit{apparent} efficiency of this conversion process: 
\bb
\eta_\text{app} = \frac{J_L}{\dot{E}_M}\,,
\ee
plotted in Fig.~\ref{f:currents}c). 
This quantity is maximized for $\gamma_M > g$ and $T_R \gtrsim T_L$ where it reaches the value $\eta_\text{app} \simeq \frac{\Delta (\mu-\epsilon)}{\Omega^2}-\frac{1}{2}$, setting a condition for refrigeration to happen on the dot parameters and the reservoir chemical potential: $\Delta (\mu-\epsilon) >\Omega^2/2$. Efficiency $\eta_\text{app}$ brings some information about the process from the standpoint of the working body dynamics, that is, when considering $\dot{E}_M$ as the resource fueling the process. This reasoning is analogous to the standard thermodynamic analysis of heat engines, where the heat from the hot bath (resp. the power) received by the working medium is identified with the ideal cost to operate the engine. However, one can wonder to which extent the energy flows $\dot{E}_M$ reflects the actual resource cost of the measurement, even under ideal conditions. Indeed, the latter, has been shown to include other contributions on top of the energy transferred to the system, related to the amount of acquired information, and entropy produced during the measurement \cite{Latune25}. In particular, treating the measuring apparatus as a black box can hide an energy balance, where a work input of magnitude larger than $\dot{E}_M$ is partly dissipated as heat inside the apparatus.
To gain further insights into the machine's efficiency and the energetics of the measurement process, we introduce in the following a thermodynamic model of the measuring apparatus.
\\

\section{General thermodynamic model of the measuring apparatus}
 In full generality, it is reasonable to assume that the measuring apparatus operates in contact with an environment at temperature $T_M$ (for instance, constituted by its internal degrees of freedom), while being powered by a work source (see Fig.~\ref{f:Setup}). The energy flow $\dot E_M$ received by the working body therefore originates from a combination of heat $J_M$ and work $P_M$ inputs received by the measuring apparatus:
\bb\label{eq:1stLawQPC}
\dot{E}_M = J_M + P_M\,.
\ee
Under this generic description, the second law of thermodynamics applied to the extended machine, including the measuring apparatus, imposes at steady state (see Fig.~\ref{f:Setup} for the sign convention):
\bb
-\frac{J_L}{T_L} - \frac{J_R}{T_R} - \frac{J_M}{T_M} \geq 0\,.
\ee
Using Eq.~\eqref{eq:first_law}, we obtain:
\bb
J_L \leq \frac{T_L}{T_R - T_L} \left( P_M + J_M \left( \frac{T_M - T_R}{T_M }\right)  \right).
\ee
Two limiting operating regimes then emerge for the extended machine. First, when $T_M = T_R$, \textit{i.e.} when the measuring apparatus operates at the hot bath, refrigeration $J_L\geq 0$ is possible only if $P_M > 0$. This regime corresponds to a conventional work-fueled refrigerator, whose coefficient of performance is limited by Carnot's bound $J_L/P_M \leq \eta_C = T_L/(T_R-T_L)$.
In contrast, when $P_M=0$ and $T_M > T_R$, refrigeration is possible provided $J_M \geq 0$, that is when the measuring apparatus provides heat to the working body. The device then behaves as an autonomous absorption refrigerator \cite{Srikhirin01,Levy11,Manikandan20,Bhandari21} with three reservoirs at different temperatures, characterized by the coefficient of performance $J_L/J_M \leq \frac{T_R^{-1}-T_M^{-1}}{T_L^{-1}-T_R^{-1}}$. In the general case, the refrigeration process can be fueled by a combination of heat (whenever $J_M\geq 0$) and work (whenever $P_M\geq 0$). The efficiency of conversion of this hybrid combination of resources into positive cooling power is quantified by \cite{Manzano20}:
\bb\label{eq:COP}  
\eta = \frac{J_L}{P_M \Theta(P_M) + J_M\left(1-\frac{T_R}{T_M}\right)\Theta(J_M)} \leq \eta_C,
\ee
where $\Theta$ denotes Heaviside's step function. Note that, the extended machine including the measuring apparatus in the machine is fueled by conventional thermodynamic resources, and it is now reasonable to identify the work and heat inputs of the apparatus, whenever they are positive, with the cost to operate the machine (in ideal conditions).

This general analysis shows that the regime of operation and the actual thermodynamic efficiency are controlled by the ratio $\xi = J_M/P_M$ between work and heat inputs of the apparatus, which in turn depends on its microscopic details.
The measurement-induced dynamics of the working body (characterized by the measurement rate $\gamma_M$) only fixes the value of the total energy flow $\dot E_M$ received by the dots.
In the following, we analyze a microscopic model of the apparatus which allows us to vary the work versus heat parts of $\dot E_M$, while keeping the measurement-induced dynamics on the system fixed.\\

\begin{figure}
\includegraphics[width=\columnwidth]{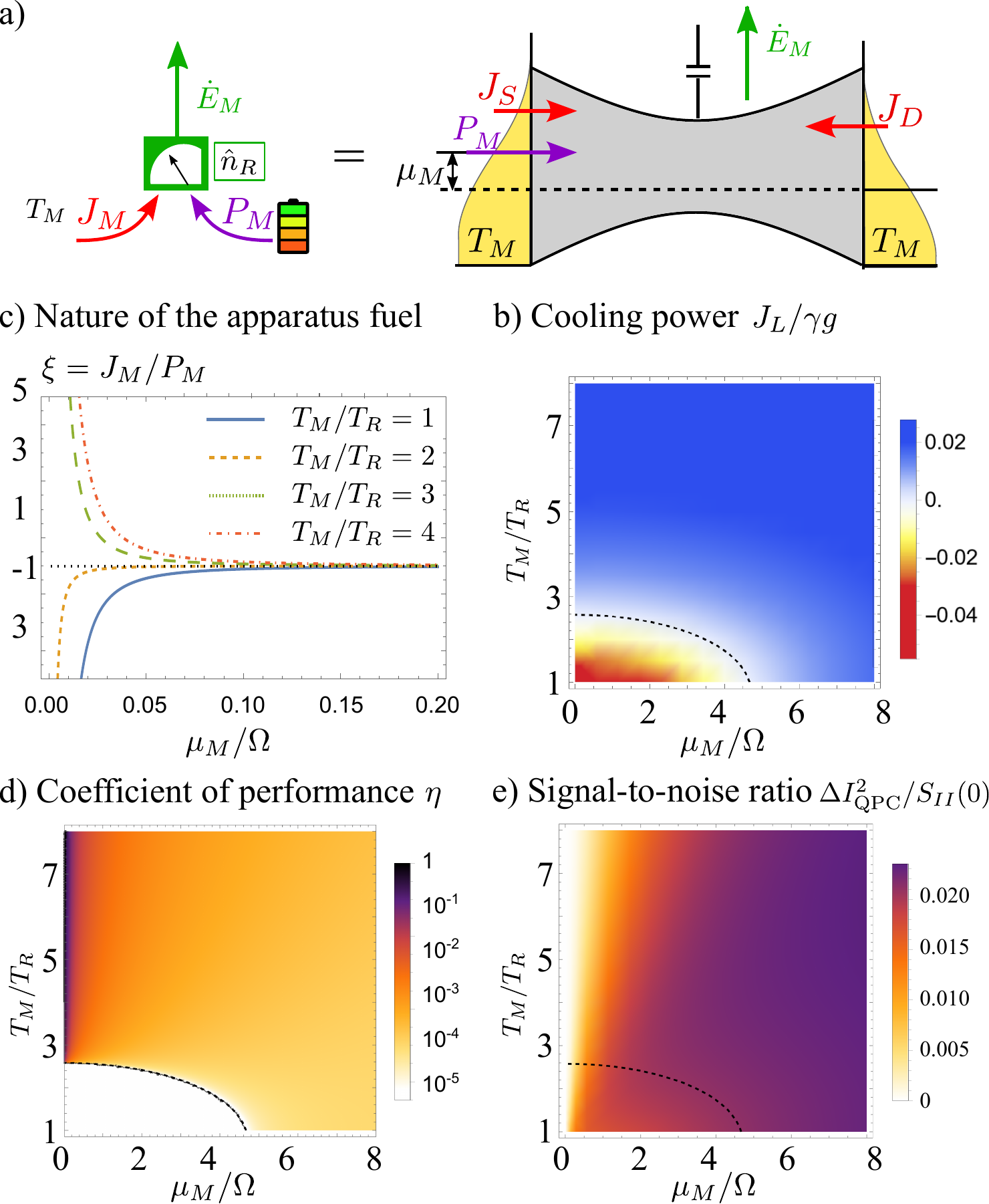}
\caption{a): Realistic measuring apparatus model based on a tunnel junction (QPC) capacitively coupled to the right QD. b): Cooling power $J_L/\gamma g$,  c): Coefficient of performance $\eta$ of the hybrid machine, defined in Eq.~\eqref{eq:COP} and d): Signal-to-noise ratio, 
 as a function of the measuring apparatus temperature $T_M$ and electric bias $\mu_M$. The dashed black line delimits the area $J_L \leq 0$ in which refrigeration is not possible. e): Ratio $\xi = J_M/P_M$ of the heat and power flows used to fuel the measuring apparatus as a function of the electric bias $\mu_M$, for different values of the measuring apparatus temperature $T_M$. The other parameters are the same as in Fig.~\label{f:QPC}}
\end{figure}

\section{QPC as a continuous charge detector}

A QPC is an electrical conductor operating here in the single channel limit whose transmission probability $\mathcal{T}(E)$ depends on the charge state of a nearby dot via capacitive coupling \cite{Korotkov99,Buttiker00,Goan01,Ihn10,Cuetara15}. Under potential bias, it provides a continuous charge measurement, the result being encoded in the current flowing through it. Hereafter, we call the leads coupled to the QPC source (S) and drain (D) to distinguish them from the $L,R$ reservoirs coupled to the working body. 
They are characterized by Fermi distributions $f_S$ and $f_D$, respectively. A schematic of this situation is shown in Fig.~\ref{f:QPC}{\bf a}. 
Due to electronic shot noise \cite{Blanter00}, the measurement takes a finite time to completely project the QD onto one of its charge state. Reciprocally, over a finite integration time $\Delta t$ (corresponding to the coarse-graining time step of the Lindblad equation), the method provides a weak measurement \cite{Jordan24}. The latter costs work associated with maintaining a voltage bias $\Delta V = \mu_M/e$ between source and drain of the QPC, leading to an electrical power consumption $P_M = \Delta V I_{QPC}$. 
The heat flow $J_M$ equals the total heat current from the
source and drain leads.

From the microscopic model corresponding to the QPC \cite{Goan01,Cuetara15}, we derive the measurement-induced Lindbladian (see Appendix \ref{s:App_QPC_dyn} for a detailed derivation):
\bb
{\cal L}_\text{QPC} = \sum_{\omega = 0,\pm\Omega} \gamma_\text{QPC}(\omega)D[\hat n_\omega],
\ee
with
\bb
\gamma_\text{QPC}(\omega) &=& \int dE \,\, {\cal T}_M(E)f_S(E)(1-f_D(E-\omega)) \nonumber\\
&&+\int dE \,\,{\cal T}_M(E)(1-f_S(E))f_D(E+\omega)\,.\label{eq:gammaM}
\ee
In the above, we have defined ${\cal T}_M(E) = (\sqrt{{\cal T}_1(E)}-\sqrt{{\cal T}_0(E)})^2$, where ${\cal T}_{0/1}(E)$ is the transparency of the QPC at energy $E$ when the right dot is empty/occupied. Importantly, the Lindbladian $\mathcal{L}_\text{QPC}$ reduces to $\mathcal{L}_M$ only when $\gamma_\text{QPC}(\omega)$ is energy-independent, \textit{i.e.} $\gamma_\text{QPC}(\omega) = \gamma_\text{QPC}(\epsilon)\equiv \gamma_M$. This holds true either when $\mu_M\gg\Omega$ (the electric bias powering the measuring apparatus is much larger than the scale of transitions induced in the working body) or when $T_M \gg T_R$ (the apparatus operates at a temperature larger than the two other heat sources, the left and right reservoirs coupled to the QDs). When using ${\cal L}_\text{QPC}$ instead of ${\cal L}_M$ in the ME \eqref{eq:ME}, we find that the refrigerator progressively stops working when we exit this ideal regime ($\gamma_\text{QPC}(\omega)=\gamma_M$).  Figure~\ref{f:QPC} b) shows the cooling power density plot. $J_L$ becomes negative within the region delimited by the dotted line, which corresponds to the regime where $\gamma_\text{QPC}$ is strongly frequency-dependent.

From our microscopic model, we also compute the steady-state electric current through the QPC $I_\text{QPC}$, electric power $P_M$, and heat flow $J_M$ (see Appendix \ref{s:App_QPC_energy} for the derivation and expressions) to assess the ratio $\xi=J_M/P_M$ as a function of the QPC parameters and the coefficient of performance $\eta$ (see Eq.~ \eqref{eq:COP}). Results are shown in Fig.~\ref{f:QPC}c) and d). We observe that our model can operate in both the work-fueled and heat fueled regime depending on the QPC parameters.

The first regime occurs when $\mu_M \ll\Omega$, and the ratio $\xi$ diverges. It corresponds to the machine being mainly fueled by heat coming from the QPC leads. Refrigeration ($\xi > 0$) then requires a sufficiently high QPC temperature $T_M>T_R$ and, as discussed earlier, the machine operates as an absorption fridge. The measuring apparatus transfers almost all the heat provided by its reservoirs to the working body $\dot{E}_M \simeq J_M$.
Interestingly, this regime achieves high values of the coefficient of performance, almost reaching the apparent efficiency, $\eta \simeq \eta_\text{app}$. Indeed, in this regime, almost all the energy provided by the measuring apparatus bath is funneled to the dots.

The second regime corresponds to $\mu_M\gtrsim \Omega$, when the ratio $\xi$ goes to $-1$. In this case, electrical work is consumed by the apparatus, and is almost entirely dissipated as heat into the QPC reservoirs, $J_M \simeq -P_M$. Only a small fraction of it is transferred to the working body and fuels the refrigerator, $\dot{E}_M \ll P_M$. The machine then works as a conventional refrigerator, with low overall coefficient of performance (see Fig.~\ref{f:QPC}
d).

Even when $\gamma_\text{QPC}$ is frequency-independent, the measurement performed by the QPC remains nonideal. To relate the thermodynamic performance of the refrigerator to the efficiency of the measurement which fuels it, we compute the signal-to-noise ratio (SNR) of the QPC charge current 
\bb
r = \Delta I_{QPC}^2/S_{II}(0).
\ee
The expression of $r$ and its derivation are provided in App.~\ref{s:App_QPC_SNR}.
As shown in Fig.~\ref{f:QPC} e), higher SNRs, indicating more efficient charge measurements are only obtained in the second regime when $\mu_M \gtrsim \Omega$  (energetically wasteful regime). In contrast, the heat-fueled regime leads to a vanishing SNR. This observation is in full agreement with the analysis of \cite{Latune25} where (i) the work cost of measurement was shown to grow with the amount of extracted information, and (ii) the possibility for measurement-fueled engine to achieve better thermodynamic performance when using inefficient (noisy) measurements was predicted. Our measurement-fueled refrigerator is a paradigmatic example verifying these predictions. We stress that an important property of the machine to be able to exploit inefficient measurements is the absence of any feedback stroke, unlike some other measurement-powered engines \cite{Elouard17,Elouard18,Bresque21,Fadler23,Dassonneville2025}.\\

\section{Conclusion}
 We analyzed a measurement-based quantum refrigerator generating steady-state cooling power, whose fuel is brought by the continuous detection of a local electric charge in a double quantum dot system. By introducing a general thermodynamic model for the measuring apparatus, we have shown that the machine can be either fueled by heat (as an absorption fridge), fueled by work (as a conventional fridge), or be an hybrid of both. Without a microscopic model of the measuring apparatus, it is not possible to distinguish between the different operating modes. 
We have therefore considered a realistic implementation of such a measuring apparatus, in the form of a QPC. The machine can be tuned to be heat-fueled or work-fueled by changing the microscopic parameters of the model (potential bias and temperature). 
Our analysis shows a trade-off between the thermodynamic efficiency of the refrigerator, and the measurement efficiency, shedding a new light on the thermodynamic balance behind a quantum measurement, and revealing how microscopic parameters impact the nature of measurement-induced energy transfers.
While ideal measurements mostly cost and provide work \cite{Jacobs09}, and are extremely wasteful \cite{Guryanova20}, realistic noisy measurements provide heat and can  
achieve better energy-conversion efficiency. Our results pave the roads towards energetic optimization of quantum measurement protocols, a crucial step in the development of quantum computing. An interesting follow-up would be to relate the heat content of the energy provided by the measurement to measures of the measurement efficiency accessible in the dot dynamics, in contrast with SNR. Candidates include the purity of the dot state conditioned to a given value of the QPC current \cite{Bettmann23}, or the efficiency coefficient appearing in the dot stochastic master equation \cite{Jacobs06}. Our predictions could be verified experimentally in state-of-the-art nanoeletronic setups, where the double quantum dot can be designed and locally readout by probing the charge currents across a QPC \cite{Petta04}. Steady state heat current detection has been demonstrated in similar setups, e.g. based on local thermometry \cite{Majidi22,Majidi24}.\\

\noindent\emph{Acknowledgements.--} We thank Gianmichele Blasi and Mark Mitchison for fruitful discussions. C.E. acknowledges funding from French National Research Agency (ANR) under project ANR-22-CPJ1-0029-01. The work of SKM is supported in part by the Swedish Research Council under Contract No. 335-2014-7424 and in part by the Wallenberg Initiative on Networks and Quantum Information (WINQ). G.H. thanks the support of NCCR SwissMAP through the Swiss National Science Foundation.

% \bibliography{biblio}
%\input{biblio.bbl}

\pagebreak
\widetext

\appendix

% \onecolumngrid
% \vspace{\columnsep}
% \begin{center}
% \textbf{\large Supplemental Material: Revealing the fuel of a quantum continuous measurement-based refrigerator}
% \end{center}

%%%%%%%%%% Merge with supplemental materials %%%%%%%%%%
%%%%%%%%%% Prefix a "S" to all equations, figures, tables and reset the counter %%%%%%%%%%
\setcounter{equation}{0}
\setcounter{figure}{0}
\setcounter{table}{0}
\setcounter{section}{0}
\makeatletter
\renewcommand{\theequation}{S\arabic{equation}}
\renewcommand{\thefigure}{S\arabic{figure}}
%\renewcommand{\bibnumfmt}[1]{[S#1]}
%\renewcommand{\citenumfont}[1]{S#1}
%%%%%%%%%% Prefix a "S" to all equations, figures, tables and reset the counter %%%%%%%%%%

\section{Microscopic derivation of the master equation} \label{s:App_ME}

\subsection{Coarse-graining and secular approximation}

\subsubsection{Microscopic model}

The leads $\alpha=L,R$ are microscopically described by bare Hamiltonians $H_\alpha$, and their coupling $V_\alpha$ to the dots:
\bb 
H_\alpha = \sum_k \epsilon_k d_{\alpha,k}^\dagger d_{\alpha,k}, \quad\quad\quad
V_\alpha =\sum_k g_{\alpha,k}d_{\alpha,k} c^\dagger_{\alpha} + h.c.
\ee
with $d_{\alpha,k}$ the fermionic annihilation operator for an electron in mode $k$ of reservoir $\alpha$, and $c_L = \ket{00}\bra{10}$, $c_R=\ket{00}\bra{01}$ anihilate an electron in the left and right dot, respectively. It is also useful to express the double dot Hamiltonian in terms of the local operators:
\bb 
H_\text{QD} = \epsilon\left( c_L^\dagger c_L + c_R^\dagger c_R \right) + \frac{\Delta}{2}\left(c_L^\dagger c_L - c_R^\dagger c_R\right)+\frac{g}{2}\left(c_L^\dagger c_R + c_R^\dagger c_L\right).
\ee

In the absence of measurement, the joint dots-leads density operator $\rho_\text{tot}$ obeys the evolution equation:
\bb\label{eq:VNE0}
\dot\rho = -i[H_\text{QD}+\sum_{\alpha=L,R}(H_\alpha+V_\alpha),\rho_\text{tot}]
\ee

\subsubsection{Dissipative contributions from the leads}

The Lindbladians ${\cal L}_\alpha$ induced by the leads can be derived from a time coarse-graining of the exact Liouville equation for the joint QDs-leads system with time step $\Delta t$, followed by a trace over the leads' degrees of freedom. More precisely, the dissipation terms are derived from:
\bb \label{eq:cg}
\frac{\Delta \rho^I}{\Delta t} = \text{Tr}_{L,R}\frac{1}{\Delta t}\int_{t}^{t+\Delta t} \dot \rho_\text{tot}^I(t')dt',
\ee
with superscript $I$ denoting the interaction picture with respect to $H_0 = H_\text{QD}+\sum_\alpha H_\alpha$, in which the dots-lead evolution equation is
\bb
 \dot\rho^I(t)= \frac{1}{\Delta t}\int_{t}^{t+\Delta t}\left[\sum_{\alpha=L,R}V_\alpha^I(t'),\rho_\text{tot}^I(t')\right]dt', 
 \ee
 with $X^I(t) = e^{iH_0 t}Xe^{-i H_0t}$ for any operator $X$.

Assuming that lead $\alpha$ is initially in at thermal equilibrium, and after performing Born and Markov approximations \cite{CohenBook,Breuer02} we get:
\bb
{\cal L}_\alpha\rho = -\frac{1}{\Delta t}\text{Tr}_{\alpha}\left\{\int_{t}^{t+\Delta t}ds\int_0^\infty d\tau [V_\alpha^I(s),[V_\alpha^I(s-\tau),\rho(s)\otimes\rho_\alpha^\text{eq}]]\right\}\label{prec}
\ee 
The explicit form of the Lindbladian ${\cal L}_\alpha$ is then obtained from the decomposition of $V_\alpha$ in the basis of eigenoperators of $H_\text{QD}$, and a crucial approximation called the secular approximation \cite{CohenBook}. This approximation consists in neglecting in Eq.\eqref{prec} the terms oscillating in time $s$ at frequency $\omega$ larger than $\Delta t^{-1}$. Indeed, these terms will have a smaller weight than the other terms by a factor $\mathrm{sinc}(\omega\Delta t/2)\ll 1$, with $\omega\Delta t\gg 1$. The approximation therefore requires a sufficiently long coarse-graining time, fulfilling:

\bb
\gamma, \omega_j^\text{nonsec} \ll \Delta t^{-1} \ll \omega_i^\text{sec}, \tau_\text{c}^{-1},
\ee
where $\tau_\text{c}$ is the memory time of the bath, $\gamma$ the magnitude of the dissipation-rate induced by the leads and $\{\omega_j^\text{nonsec},\omega_i^\text{sec}\}$ are frequencies involved in the system's free dynamics. It is only when all the system's frequencies are much larger than $\Gamma_\alpha$ that the secular approximation is fully justified.  The compliance of the dynamics with thermodynamic laws is then well-understood \cite{Soret22}.

Here, we have
\bb
c_{\alpha}^I(t) = e^{-i\epsilon t} \sum_{l=\pm,n=0,1}  e^{-il \Omega t/2}e^{-in U t}A_{\alpha,n,l}\,,
\ee
where
\begin{equation}
\left\{
\begin{array}{l}
A_{L,0,+} = \cos(\theta) \ket{00}\bra{+}\\
A_{L,0,-} = -\sin(\theta)\ket{00}\bra{-} \\
A_{R,0,+} = \sin(\theta)\ket{00}\bra{+} \\
A_{R,0,-} = \cos(\theta)\ket{00}\bra{-}
\end{array}
\right.
\quad\quad\quad
\left\{
\begin{array}{l}
A_{L,1,+} = \cos(\theta)\ket{-}\bra{11}\\
A_{L,1,-} = \sin(\theta)\ket{+}\bra{11}\\
A_{R,1,+} = -\sin(\theta)\ket{-}\bra{11}\\
A_{R,1,-} = \cos(\theta)\ket{+}\bra{11},
\end{array}
\right.
\end{equation}
are eigenoperators of $H_\text{QD}$, whose eigenstates fulfill:
\bb
\ket{+} &=& \cos(\theta)\ket{10} + \sin(\theta)\ket{01} \,,\\
\ket{-} &=& -\sin(\theta)\ket{10} + \cos(\theta)\ket{01}\,.
\ee 
Anticipating that we will assume $\epsilon\Delta t \gg 1$, we only keep in Eq.~\eqref{prec} the terms oscillating as a function of $s$ at frequencies lower than $\epsilon$, i.e. involving one $c_{L,R}$ and one $c_{L,R}^\dagger$ only:

\bb
{\cal L}_\alpha\rho &=& -\frac{1}{\Delta t}\int_{t}^{t+\Delta t}ds\int_0^\infty d\tau \moy{R_{\alpha}^{I}R_{\alpha}^{I\dagger}(-\tau)}\left(c^{I\dagger}_\alpha(s)c^{I}_\alpha(s-\tau)\rho - c^{I}_\alpha(s-\tau)\rho c^{I\dagger}_\alpha(s)\right) \nonumber\\
&&-\frac{1}{\Delta t}\int_{t}^{t+\Delta t}ds\int_0^\infty d\tau \moy{R_{\alpha}^{I\dagger}R_{\alpha}^{I}(-\tau)}\left(c^{I}_\alpha(s)c^{I\dagger}_\alpha(s-\tau)\rho - c^{I\dagger}_\alpha(s-\tau)\rho c^{I}_\alpha(s)\right) +\text{h.c.}
\ee 

When introducing the exact form of $c^{I}_\alpha(s-\tau)$, we obtain constant operators multiplied by an exponential factor of the form $e^{i\omega\tau}e^{i\nu s}$. The integration over $\tau$ can then be performed yielding numerical factors involving the bath spectral densities, namely:
\bb
S_{\alpha}(\omega) &=& \int_0^\infty d\tau e^{i\omega\tau} \moy{R_{\alpha}^{I}R_{\alpha}^{I\dagger}(-\tau)}\nonumber\\
&=& \sum_{k,k'} g_{\alpha,k}g_{\alpha,k'}\int_0^\infty d\tau e^{i(\omega-\omega_{k'})\tau} \moy{d_{\alpha,k}d_{\alpha,k'}^{\dagger}}
\nonumber\\
&=& \sum_{k} g_{\alpha,k}^2\left(\pi\delta(\omega-\omega_k)-i {\cal P}\frac{1}{\omega-\omega_k}\right) \left(1-f_{\alpha}(\omega_k)\right)\nonumber\\
&\equiv& \tfrac{1}{2}\Gamma_{\alpha,\downarrow}(\omega) + i \Delta_{\alpha,\downarrow}(\omega)
\ee
and
\bb
R_{\alpha}(\omega) &=& \int_0^\infty d\tau e^{i\omega\tau} \moy{R_{\alpha}^{I\dagger}R_{\alpha}^{I}(-\tau)}\nonumber\\
&=& \sum_{k,k'} g_{\alpha,k}g_{\alpha,k'}\int_0^\infty d\tau e^{i(\omega+\omega_{k'})\tau} \moy{d^\dagger_{\alpha,k}d_{\alpha,k'}}
\nonumber\\
&=& \sum_{k} g_{\alpha,k}^2\left(\pi\delta(\omega+\omega_k)-i {\cal P}\frac{1}{\omega+\omega_k}\right) f_{\alpha}(\omega_k)\nonumber\\
&\equiv& \tfrac{1}{2}\Gamma_{\alpha,\uparrow}(-\omega) - i \Delta_{\alpha,\uparrow}(-\omega)\,.
\ee

We have identified the rates characterizing the exchange of excitations with the bath $\Gamma_{\alpha l \uparrow} =\gamma f_{\alpha l}$ and $\Gamma_{\alpha l,\downarrow} =\gamma \left(1-f_{\alpha l})\right)$, with $f_{\alpha l} = [ e^{(\epsilon+l\Omega/2-\mu_\alpha)/kT_\alpha}+1]^{-1}$ the Fermi function of lead $\alpha$ in terms of the chemical potential $\mu_\alpha$ and temperature $T_\alpha$, and  $\gamma =  2\pi\sum_{k} g_{\alpha,k}^2\delta(\epsilon-\omega_k) \ll \epsilon$  the bare dissipation rate, introduced above, that we assume equal for both reservoirs for the sake of simplicity.

We therefore have

\bb
{\cal L}_\alpha\rho &=& -\frac{1}{\Delta t}\int_{t}^{t+\Delta t}\!\!\!\!\!\! ds \sum_{l,l'=\pm}\sum_{n,n'=0,1}  \left(A_{\alpha,n,l}^\dagger A_{\alpha,n',l'}\rho - A_{\alpha,n',l'}\rho A_{\alpha,n,l}^\dagger\right) S_{\alpha}(\epsilon+l'\Omega/2+n'U)e^{i(l-l')\Omega s/2}e^{i(n-n')U s}\nonumber\\
&&-\frac{1}{\Delta t}\int_{t}^{t+\Delta t}\!\!\!\!\!\! ds \sum_{l,l'=\pm} \sum_{n,n'=0,1} \left(A_{\alpha,n,l} A_{\alpha,n',l'}^\dagger\rho - A_{\alpha,n',l'}^\dagger\right)\rho A_{\alpha,n,l} R_{\alpha}(-\epsilon-l'\Omega/2-n'U)e^{-i(l-l')\Omega s/2}e^{-i(n-n')U s}\nonumber\\&&+\text{h.c.}\label{eq:prec}
\ee 

The integration over $s$ yields numerical coefficients of the form
\bb
\int_{t}^{t+\Delta t}\!\!\!\!\!\! ds e^{i \omega_{l,l',n,n'} s} = \Delta t e^{i \omega_{l,l',n,n'} (t+\Delta t/2)} \text{sinc}(\omega_{l,l',n,n'}\Delta t/2),
\ee
where $\omega_{l,l',n,n'} = (l-l')\Omega/2 + (n-n')U$ and $\text{sinc}(x) = \sin(x)/x$ (or the complex conjugate of the above expression). Now, we see that depending on the relationship between $\Delta t$, $\Omega$ and $U$, the $\text{sinc}$ function will be either of the order of unity $\omega_{l,l',n,n'}\Delta t \gtrsim 1$ or vanishingly small $\omega_{l,l',n,n'}\Delta t \gg 1$, and the master equation will involve terms sensitive to different subsets of the frequencies $\{\omega_{l,l',n,n'}\}$, which are nothing but the Bohr frequencies of the two dot system (minus the charging energy $\epsilon$ assumed to be big). Finally, the inequality $\Delta t U \gg 1$ (resp. $\Delta t \Omega \gg 1$) imposes that only the terms $n=n'$ (resp. $l=l'$) of Eq.~\eqref{prec} must be kept in the final master equation. An even more general treatment would add another index $k = \pm$ for the dependence in $\epsilon$, define frequencies $\omega_{l,l',n,n',k,k'} = \omega_{l,l',n,n'} + (k-k')\epsilon$  allowing to study the case $\epsilon\Delta t \ll 1$ that we have ruled out from the beginning. We distinguish four cases from the most ``secular'' (all transition frequencies are resolved separately by the bath, the Lindblad operators appearing in the Lindbladian act on the total dot system, that is, are ``global'') to the most ``local'' (only $\epsilon$ is resolved, the Lindblad operators are local):
\begin{enumerate}
\item If $\epsilon,U,\Omega \gg \Delta t^{-1} \gg \Gamma_\alpha$, only the terms $l=l'$ and $n=n'$ survive.
\item If $\epsilon,U \gg \Delta t^{-1} \gg \Omega,\Gamma_\alpha$, only the terms $n=n'$ survive.
\item If $\epsilon,\Omega \gg \Delta t^{-1} \gg U,\Gamma_\alpha$, only the terms $l=l'$ survive.
\item If $\epsilon \gg \Delta t^{-1} \gg U,\Omega,\Gamma_\alpha$, all the terms survive.
\end{enumerate}

After the corresponding terms have been suppressed, the final master equation is obtained by going back to the Schr\"odinger picture, removing any remaining time dependence. The regime suitable for the measurement-based refrigerator corresponds to case 1, which is also called global regime as it features incoherent transitions between the global eigenstates of the two-dot system.

\subsubsection{Contribution from the measurement: phenomenological treatment}

We first treat phenomenologically the continuous measurement of the right-dot occupation $\hat n_R$. A sequence of weak (incomplete) measurements \cite{WisemanBook}, that is each associated with a short interaction time-scale with the system $\tau_\text{meas}$ leads to an average dynamics of the form \cite{Jacobs06}
\bb 
\dot{\rho}\Big\vert_\text{meas} := \frac{\rho(t+\tau_\text{meas})-\rho(t)}{\tau_\text{meas}}\simeq  {\cal L}_\text{meas}\rho = \gamma_M {\cal D}[\hat n_R]\rho.
\ee
Above, $\gamma_M$ is the measurement strength, verifying $\gamma_M\tau_\text{meas}\ll 1$ such that the measurement is complete (and the coherences in the eigenbais of $\hat n_R$ have vanished) after a time $t > \gamma_M$.\\

Assuming the ideal case where $\tau_\text{meas}$ is the shortest time scale of the problem, the contribution ${\cal L}_\text{meas}\rho$ can be directly added to the evolution equation \eqref{eq:VNE0} for $\dot\rho_\text{tot}$. This term may be affected by the coarse-graining as the measured observable $\hat n_R$ does not commute with Hamiltonian $H_R$. Assuming $\gamma_M\Delta t \ll 1$,the coarse-grained contribution of the continuous measurement reads:
\bb\label{eq:LMdef}
{\cal L}_M^I(t)\rho^I = \frac{1}{\Delta t} \int_{t}^{t+\Delta t} \gamma_M {\cal D}[\hat n_R^I(t')]\rho^I(t')dt'\simeq \frac{1}{\Delta t} \int_{t}^{t+\Delta t} \gamma_M {\cal D}[\hat n_R^I(t')]dt'\rho^I(t),
\ee
where we have used that in the interaction picture, the typical evolution timescale of the density operator is $\gamma^{-1}\gg \Delta t$.

\subsection{Global regime}

In the main text, we focus on the global regime, corresponding to sufficiently strong interdot interaction, allowing for a coarse-graining time-scale $\Delta t \gg \Omega^{-1}$. More precisely, we require:
\bb 
 \epsilon, U, \Omega \gg \Delta t^{-1} \gg \gamma.
\ee

\subsection{Dissipation due to the leads}
In this case we only keep the terms $n =n'$ and $l=l'$, leading to:
\bb
{\cal L}_\alpha\rho &=&  -\sum_{l=\pm}\sum_{n=0,1}\left(A_{\alpha,n,l}^\dagger A_{\alpha,n,l}\rho - A_{\alpha,n,l}\rho A_{\alpha,n,l}^\dagger\right) S_{\alpha}(\epsilon+l\Omega/2+n U)\nonumber\\
&&-\sum_{l=\pm}\sum_{n=0,1}  \left(A_{\alpha,n,l} A_{\alpha,n,l}^\dagger\rho - A_{\alpha,n,l}^\dagger\rho A_{\alpha,n,l}\right) R_{\alpha}(-\epsilon-l\Omega/2-n U)+\text{h.c.}\nonumber\\
&=&-i[H_{\alpha}^{(LS)},\rho]+\sum_{l=\pm}\sum_{n=0,1} (  \Gamma_{\alpha,\downarrow}(\epsilon+l\Omega/2+n U) D[ A_{\alpha,n,l}]+  \Gamma_{\alpha,\uparrow}(-\epsilon-l\Omega/2-n U) D[ A_{\alpha,n,l}^\dagger])\,,
\label{eq:GlobLimb}
\ee 
where 
\bb  
H_{\alpha}^{(LS)} = \sum_{l=\pm}\sum_{n=0,1} \lambda_l^2\left[\Delta_{\alpha,\downarrow}(\epsilon+\frac{l\Omega}2+nU)A^\dagger_{\alpha,n,l}A_{\alpha,n,l}+\Delta_{\alpha,\uparrow}(-\epsilon-\frac{l\Omega}2-nU)A_{\alpha,n,l}A^\dagger_{\alpha,n,l}\right]\label{eq:HLS}
\ee
is a small renormalization of the dot energy transition due to the coupling to the reservoirs (Lamb shift) \cite{Breuer02}. 
Finally, we assume that $U\gg T_\alpha$ such that the two-electron state $\ket{11}$ is negligibly populated (single-electron regime) and we may keep only the terms $n=0$ in Eq.~\eqref{eq:GlobLimb}. Moreover, we rewrite
\bb 
\Gamma_{\alpha,\uparrow}(-\epsilon-l\Omega/2) = \Gamma_{\alpha,l}f_{\alpha,l}, \quad \Gamma_{\alpha,\downarrow}(-\epsilon-l\Omega/2) = \Gamma_{\alpha,l}(1-f_{\alpha,l})\rho,
\ee
with
\bb 
\Gamma_{\alpha,l}= 2\pi\sum_k g^2_{\alpha,k}\delta(\epsilon+l\Omega/2-\omega_k),\quad
f_{\alpha,l}=[e^{(\epsilon+l\Omega/2-\mu_\alpha)/kT_\alpha}+1]^{-1},
\ee
and assume that the coupling strength $g_{\alpha,k}$ are identical for both leads (do not depend on $\alpha$) and depend slowly enough in $\omega_k$ to approximate $\Gamma_{\alpha,l}\simeq 2\pi\sum_k g^2_{\alpha,k}\delta(\epsilon-\omega_k)= \gamma$, $\forall l$, leading to Eq.~\eqref{eq:LLR} of the main text once back in the Schr\"odinger picture.

\subsubsection{Contribution of the continuous measurement}

As the charge operator in the interaction picture $\hat{n}^I_R(t)  = \sum_{\omega=0,\pm\Omega} \hat n_\omega e^{i\omega t}$ evolves over a time-scale $\Omega^{-1}$, the continuous measurement contribution Eq.~\eqref{eq:LMdef} is strongly affected by the coarse-graining over $\Delta t \gg \Omega^{-1}$. All contributions 
oscillating at frequency $\Omega$ average out, leading to Eq.~\eqref{eq:LM} of the main text in the Schr\"odinger picture.

\subsection{Local regime}

\subsubsection{Master equation}

 We have also investigated the machine in case 2 defined above, where only the $n \neq n'$ can be neglected, while all the terms in the sum over $l,l'$ must be kept. After writing the result of this procedure back in the Schr\"odinger picture, we obtain:

\bb\label{eq:Lalphanoloc}
{\cal L}_\alpha\rho &=&  -\sum_{n=0,1}\sum_{l,l'=\pm}\left(A_{\alpha,n,l}^\dagger A_{\alpha,n,l'}\rho - A_{\alpha,n,l'}\rho A_{\alpha,n,l}^\dagger\right) S_{\alpha}(\epsilon+l'\Omega/2+n U)\nonumber\\
&&-\sum_{n=0,1}\sum_{l,l'=\pm} \left(A_{\alpha,n,l} A_{\alpha,n,l'}^\dagger\rho - A_{\alpha,n,l'}^\dagger\rho A_{\alpha,n,l}\right) R_{\alpha}(-\epsilon-l'\Omega/2-n U)+\text{h.c.}
\ee 
Note that we have performed the integration over $s$ whose only effect is to cancel the $1/\Delta t$ factor. We now apply the assumption that the reservoir's spectral density is flat over the frequency scale $\Omega$, \textit{i.e.} we assume the following substitutions are valid: 
\bb
S_{\alpha}(\epsilon + (l-l')\Omega/2 + nU) &\to&  S_{\alpha}(\epsilon+ nU)\\
R_{\alpha}(-\epsilon - (l-l')\Omega/2-n U) &\to& R_{\alpha}(-\epsilon-n U)\,.\label{eq:LocalApprox}
\ee
This approximation is expected to be valid in the regime $\Omega \ll T_{\alpha}$, and for a sufficiently slow-varying spectral density of the leads. We also use the identity:
\bb
\sum_{l} A_{\alpha,n,l}  e^{-il\Omega s/2} = \ket{0}_\alpha\bra{1}\otimes \ket{n}_{\bar\alpha}\bra{n}\,,
\ee
and, as before, apply the single-electron approximation (neglecting population of state $\ket{11}$), and neglect a Hamiltonian's correction to $H_{QD}$ analogous to $H_\alpha^{(LS)}$. We obtain:
\bb
{\cal L}_\alpha^\text{(loc)}\rho&=& \Gamma_{\alpha,\downarrow}(\epsilon)D[c_\alpha] +\Gamma_{\alpha,\uparrow}(\epsilon)D[c_\alpha^\dagger]\,.\label{eq:Lalphaloc}
\ee  

We can see that each lead $L$ now induces a ``local'' Lindbladian, acting only onto the dot it is coupled to, as in the absence of inter-dot coupling. In the same regime, the coarse-graining time is much shorter than the evolution time-scale $\Omega^{-1}$ of $\hat n^I_R(t)$, such that the phenomenological measurement-induced Lindbladian takes the form (back in the Schr\"odinger picture):
\bb 
{\cal L}_M^\text{(loc)}= {\cal L}_\text{meas} = \gamma_M D[\hat n_R].
\ee

\subsubsection{Error due to the local approximation}

To test the stability of the local approximation in our system, we introduce the difference between Eq.~\eqref{eq:Lalphanoloc} (before the approximation) and Eq.~\eqref{eq:Lalphanoloc} (after), in the single-electron regime:
\bb 
\delta {\cal L}_\alpha = {\cal L}_\alpha-{\cal L}_\alpha^\text{(loc)} &=&  -\sum_{l,l'=\pm}\left(A_{\alpha,0,l}^\dagger A_{\alpha,0,l'}\rho - A_{\alpha,0,l'}\rho A_{\alpha,0,l}^\dagger\right) \delta S_{\alpha,l'}\nonumber\\
&&-\sum_{l,l'=\pm} \left(A_{\alpha,0,l} A_{\alpha,0,l'}^\dagger\rho - A_{\alpha,0,l'}^\dagger\rho A_{\alpha,0,l}\right) \delta R_{\alpha,l'}+\text{h.c.}\,,
\ee
with $\delta S_{\alpha,l'} = S_{\alpha}(\epsilon+l'\Omega/2+n U)-S_\alpha(\epsilon+n U)$ and
$\delta R_{\alpha,l'}=R_{\alpha}(-\epsilon-l'\Omega/2-n U)-R_{\alpha}(-\epsilon-n U)$.
Neglecting the Hamiltonian part, we obtain:
\bb 
\delta {\cal L}_\alpha  &=&  \sum_{l,l'=\pm}\left(A_{\alpha,0,l}^\dagger A_{\alpha,0,l'}\rho - A_{\alpha,0,l'}\rho A_{\alpha,0,l}^\dagger\right)\Gamma_\alpha\delta f_{\alpha,l'} \nonumber\\
&&-\sum_{l,l'=\pm} \left(A_{\alpha,0,l} A_{\alpha,0,l'}^\dagger\rho - A_{\alpha,0,l'}^\dagger\rho A_{\alpha,0,l}\right) \Gamma_\alpha\delta f_{\alpha,l'}+\text{h.c.}\,,
\ee
with $\delta f_{\alpha,l'}=f_{\alpha}\left(\epsilon+l'\frac{\Omega}{2}\right)-f_{\alpha}\left(\epsilon\right) =-\frac{l'\Omega}{2kT_\alpha}f_{\alpha}\left(\epsilon\right)(1-f_{\alpha}\left(\epsilon\right)) + {\cal O}(\frac{\Omega}{kT_\alpha})^2$.

\subsubsection{Energy flows in the local regime}

After solving the master equation in the local regime in the steady state, we inject the solution into the definitions of the heat provided by the leads and the energy flow from the measuring apparatus, and we get:

\bb
J_L &=& A^{-1} (f_R-f_L)g^2 j_+,\quad\quad
J_R = A^{-1} (f_R-f_L)g^2 j_-,\quad\quad
\dot{E}_M = A^{-1} (f_R-f_L)g^2 \gamma_\text{M} \Delta,
\ee
where we have defined a positive constant $A$ verifying
\bb
A = \frac{g^2(\gamma_M+\Gamma_\downarrow)(\Gamma_\downarrow+2\Gamma_\uparrow)}{\Gamma_L\Gamma_R} +(1-f_L f_R)((\gamma_M+\Gamma_\downarrow)^2+4\Delta^2)
\ee
and energy currents
\bb
 j_\pm = \pm \gamma_M (\epsilon_\pm -\mu) \pm\Big(\Gamma_{L,\downarrow}(\epsilon_--\mu) +\Gamma_{R,\downarrow}(\epsilon_+-\mu)\Big),\quad\quad \label{eq:jpm}
\ee
where $\epsilon_\pm = \epsilon\pm \Delta/2$ are the local energies of the left and right dots respectively. We have introduced the short-hand notation $\sum_{\alpha = L,R} \Gamma_{\alpha,\uparrow} = \Gamma_\uparrow$ (and similarly for $\Gamma_\downarrow$). These expressions verify as expected: $ J_L +  J_R +  J_M = 0 = dU/dt$. 

We can see that all the flows are proportional to $g^2$, which means that they vanish in the absence of inter-dot coupling. They can be tracked back to the contribution of the dot internal energy associated with the inter-dot coupling $V_{L-R}=\frac{g}{2}\left(c_L^\dagger c_R + c_R^\dagger c_L\right)$. This scaling makes these steady-state energy flows of the same order of magnitude or smaller than the errors due to the local approximation, which we confirmed by a numerical analysis.

\subsubsection{Conditions for refrigeration}

Refrigeration occurs when $J_L\geq 0$, i.e. when $(f_R-f_L) j_+ > 0$, which implies a sufficiently large measurement rate:
\bb 
\gamma_M \geq  \frac{\Delta(\Gamma_R f_R-\Gamma_L f_L) - 2(\epsilon-\mu)\Gamma_\downarrow}{\Delta + 2(\epsilon-\mu)}.
\ee
 In addition, $\Delta$ and $(\epsilon-\mu)$ must have opposite signs. When this holds, we find that the energy flow exchanged with the measuring apparatus has the sign of $\Delta (f_R-f_L)$, which is negative. This is surprising at it implies that the measurement process takes away energy while both the hot and cold reservoirs are cooled down. This seems to be in contradiction with the fact that continuous unread measurements of a quantum system always increase its entropy. This results casts additional doubts on the validity of the heat flows predicted by the local Lindblad equation for this problem. See also \cite{Bettmann23} for issues with the local master equation approach in a similar model used as an engine rather than a refrigerator. \\

 The results of the last two sections suggest that, in the weak lead-QDs coupling regime, the machine can generate significant cooling power only in the global regime where $\Omega = \sqrt{g^2+\Delta^2}\gg \Gamma_{L,R}$.

 \section{Measurement performed owing to a quantum point contact (QPC)} \label{s:App_QPC}
 
 \subsection{Microscopic model of the QPC}
We use the model presented in \cite{Goan01}. The total Hamiltonian is $H = H_0 + V$ with $H_0 = H_\text{QD} + H_S + H_D$ and:
\bb
H_S &=& \sum_q \omega^S_q s_q^\dagger s_q, \quad\quad
H_D = \sum_p \omega^S_p d_p^\dagger d_p\quad\quad
V = \sum_{qp} (g_{qp}+\chi_{qp} \hat n_R)(s_q^\dagger d_p + d_p^\dagger s_q) = \Upsilon + \Xi \hat n_R + \text{H.c.}\,.
\ee
Here $s_q$ and $d_p$ are fermionic annihilation operators associated with QPC source and drain electron reservoirs. We have introduced the projectors $\Pi_\pm$ acting onto the dot eigenstates $\ket{\pm}$,
such that the occupation of the right dot verifies $\hat n_R = \ket{01}\bra{01} = \sin(\theta)^2 \Pi_{+}  +\cos(\theta)^2\Pi_{-} +  \cos(\theta)\sin(\theta)(\ket{+}\bra{-}+\ket{-}\bra{+})$. In the interaction picture with respect to Hamiltonian $H_0$, the density operator $\rho_\text{tot}^I$ of the QDs and QPC evolves according to:
\bb
\dot \rho_\text{tot}^I = -i [V^I(t),\rho_\text{tot}^I],
\ee
where the interaction-picture coupling Hamiltonian takes the form $V^I(t) = \Upsilon(t) + \Xi(t) \hat n_R(t)$ with
\bb
\Upsilon(t) &=& \sum_{qp} g_{qp} s_q^\dagger d_p e^{i(\omega^S_q-\omega^D_p)t}+\text{h.c.},\quad\quad
\Xi(t) = \sum_{qp} \chi_{qp} s_q^\dagger d_p e^{i(\omega^S_q-\omega^D_p)t}+\text{h.c.},\quad\quad
\hat n_R(t) = \sum_{\omega = 0,\pm\Omega} n_\omega e^{i\omega t}\,.
\ee
Here, we have introduced the operators $\hat n_0 = \cos(\theta)^2 \Pi_{+}  +\sin(\theta)^2\Pi_{-}$, $\hat n_\Omega =  \cos(\theta)\sin(\theta)\ket{+}\bra{-}$ and $\hat n_{-\Omega} =   \cos(\theta)\sin(\theta)\ket{-}\bra{+}$. By coarse-graining the evolution equation over a time $\Delta t$, we obtain:
\bb\label{eq:VNEQPC}
\Delta \rho_\text{tot}^I = -i\int_{t}^{t+\Delta t}dt' [V^I(t'),\rho_\text{tot}^I(t)]  -\int_{t}^{t+\Delta t} dt' \int_{t}^{t'} dt''[V^I(t'),[V^I(t''),\rho_\text{tot}^I(t'')].
\ee

To express the final expressions as integrals over the energy, it is convenient to take the continuous limit of the spectra of the source and drain reservoirs. We therefore define:
\bb
{\cal T}_0(E) &=& 2\sum_{qk} g_{qk}^2 \delta(\omega_q^S-E)\nonumber\\
{\cal T}_1(E) &=& 2\sum_{qk} (g_{qk}+\chi_{qk})^2 \delta(\omega_q^S-E)\nonumber\\
{\cal T}_M(E) &=& 
2\sum_{qk} \chi_{qk}^2 \delta(\omega_q^S-E) = \left(\sqrt{{\cal T}_0(E)}-\sqrt{{\cal T}_1(E)}\right)^2,\label{d:T01M}
\ee
with $\chi_{qk}\leq 0$ (see also \cite{Goan01}).

\subsection{QPC-induced dynamics}\label{s:App_QPC_dyn}

Taking the partial trace over the source and drain spaces of Eq.~\eqref{eq:VNEQPC}, we obtain:
\bb
{\cal L}_\text{QPC}\rho &=& \frac{1}{\Delta t}\text{Tr}_\text{S,D}\{\Delta \rho_\text{tot}^I(t)\}\nonumber\\
&=&  -\frac{1}{\Delta t}\int_{t}^{t+\Delta t} dt' \int_{t}^{t'} dt''\text{Tr}_\text{S,D}\{[V^I(t'),[V^I(t''),\rho_\text{tot}^I(t'')]\}\nonumber\\
&=& -\frac{1}{\Delta t}\int_{t}^{t+\Delta t}\!\! dt' \int_{0}^{t'-t} \!\!d\tau \!\!\!\sum_{\omega,\omega'0,\pm\Omega} \!\!\!\text{Tr}_\text{S,D}\{[\Upsilon+\Xi \hat n_\omega^\dagger e^{-i\omega t'} ,[\Upsilon(-\tau)+\Xi(-\tau) \hat n_{\omega'} e^{i\omega' (t'-\tau)},\rho_\text{qd}^I(t)\otimes\rho_\text{S,D}^\text{eq}]\}\quad
\ee
We now apply the secular approximation, assuming $\Omega \gg \Delta t^{-1} \gg \gamma_M$, similar to the procedure we followed to derive the reservoir-induced dynamics in the global regime . We therefore only keep the terms $\omega=\omega'$:
\bb
{\cal L}_\text{QPC}\rho
&=& -\frac{1}{\Delta t}\int_{t}^{t+\Delta t}\!\! dt' \int_{0}^{t'-t} \!\!d\tau \text{Tr}_\text{S,D}\{[\Upsilon+\Xi \hat n_0 ,[\Upsilon(-\tau)+\Xi(-\tau) \hat n_0,\rho_\text{qd}^I(t)\otimes\rho_\text{S,D}^\text{eq}]\}\nonumber\\
&& -\sum_{\omega= \pm\Omega} \frac{1}{\Delta t}\int_{t}^{t+\Delta t}\!\! dt' \int_{0}^{t'-t} \!\!d\tau \text{Tr}_\text{S,D}\{[\Xi \hat n_\omega^\dagger ,[\Xi(-\tau) \hat n_\omega e^{-i\omega\tau},\rho_\text{qd}^I(t)\otimes\rho_\text{S,D}^\text{eq}]\}\nonumber\\
&=& \frac{1}{\Delta t}\int_{t}^{t+\Delta t}\!\! dt' \int_{0}^{t'-t} \!\!d\tau \text{Tr}_\text{S,D}\{[(\Upsilon(-\tau)+\Xi(-\tau) \hat n_0)\rho_\text{qd}^I(t)\otimes\rho_\text{S,D}^\text{eq},(\Upsilon+\Xi\hat  n_0)]\}+\text{h.c.}\nonumber\\
&& +\sum_{\omega= \pm\Omega} \frac{1}{\Delta t}\int_{t}^{t+\Delta t}\!\! dt' \int_{0}^{t'-t} \!\!d\tau e^{-i\omega\tau} \text{Tr}_\text{S,D}\{[\Xi(-\tau) \hat n_\omega  \rho_\text{qd}^I(t)\otimes\rho_\text{S,D}^\text{eq} ,\Xi \hat n_\omega^\dagger]\}+\text{h.c.}\nonumber\\
&=& \frac{1}{\Delta t}\int_{t}^{t+\Delta t}\!\! dt' \int_{0}^{t'-t} \!\!d\tau \bigg(\moy{\Xi\Xi(-\tau)}[\hat n_0\rho_\text{qd}^I(t),n_0]+\moy{\Upsilon\Xi(-\tau)}[\hat n_0,\rho_\text{qd}^I(t)]\bigg)+\text{h.c.}\nonumber\\
&& +\sum_{\omega= \pm\Omega} \frac{1}{\Delta t}\int_{t}^{t+\Delta t}\!\! dt' \int_{0}^{t'-t} \!\!d\tau \moy{\Xi\Xi(-\tau)} e^{-i\omega\tau} [\hat n_\omega\rho_\text{qd}^I(t),\hat n_\omega^\dagger]+\text{h.c.}\,.
\ee
We now use that
\bb
\text{Tr}\Big\{ \Xi \Xi(-\tau)\rho^\text{eq}_{S,D}\Big\} &=& \sum_{jk} \sum_{lm} \chi_{jk}\chi_{lm} \left(\moy{s_j^\dagger d_k s_l d_m^\dagger }e^{i(\omega^S_l-\omega^D_m)\tau}+\moy{ s_j d_k^\dagger s_l^\dagger d_m} e^{-i(\omega^S_l-\omega^D_m)\tau}\right)\nonumber\\
&=& \sum_{jk} \sum_{lm} \chi_{jk}\chi_{lm} \left(\delta_{jl}\delta_{km} \moy{s_l^\dagger s_l}\moy{d_k d_k^\dagger} e^{i(\omega^S_l-\omega^D_m)\tau}+\delta_{jl}\delta_{km} \moy{s_l s_l^\dagger}\moy{d_k^\dagger d_k} e^{-i(\omega^S_l-\omega^D_m)\tau}\right)\nonumber\\
&=& \sum_{kl}\chi_{lk}^2 \left( f^S_l(1-f^D_k) e^{i(\omega^S_l-\omega^D_k)\tau}+(1-f^S_l)f^D_k e^{-i(\omega^S_l-\omega^D_k)\tau}\right)
\ee
and similarly for $\Upsilon$ and cross-correlations between $\Xi$ and $\Upsilon$, and define
\bb
 A_{\Delta t}(u) &=& \frac{1}{u^2\Delta t}(1+iu\Delta t-e^{iu\Delta t})\\
 G_{\Delta t}(u) &=& \text{Re}A_{\Delta t}(u) =  \frac{1}{u^2\Delta t}(1-\cos(u\Delta t))\\
 F_{\Delta t}(u) &=& \text{Im}A_{\Delta t}(u) = \frac{1}{u}\left(1-\frac{\sin(u\Delta t)}{u\Delta t}\right)\,.
\ee 
We obtain:
\bb
{\cal L}_\text{QPC}\rho &=& \sum_{\omega=0,\pm\Omega}\sum_{kl}\chi_{kl}^2\bigg(A_{\Delta t}(\omega_l^S-\omega_k^D-\omega)f^S_l(1-f^D_k)+A_{\Delta t}(\omega_k^D-\omega_l^S-\omega)(1-f^S_l)f^D_k) \bigg)[\hat n_\omega\rho_\text{qd}^I(t), \hat n_\omega^\dagger]+\text{h.c.}\nonumber\\
&& +\sum_{kl}\chi_{kl}g_{kl}\bigg(A_{\Delta t}(\omega_l^S-\omega_k^D)f^S_l(1-f^D_k)+A_{\Delta t}(\omega_k^D-\omega_l^S)(1-f^S_l)f^D_k) \bigg)[\hat n_0,\rho_\text{qd}^I(t)]+\text{h.c.}\,.
\ee
We now assume that the width of the function $G_{\Delta t}(u)$, which is given by $1/\Delta t$, is negligible with respect to the typical variation scale of $f^S_l$ and $f^D_k$, which is a valid approximation when $\Delta t \gg \beta_S,\beta_D$. The regime we consider is then:
\bb
 \Delta t \gg\Omega^{-1} , T_S^{-1}, T_D^{-1}.
\ee
In this regime, we can replace the function $G_{\Delta t}$ by a Dirac distribution. Introducing:
\bb 
\gamma_\text{QPC}(\omega) =  2\sum_{kl}\chi_{kl}^2(f_l^S(1-f_k^D)\delta(\omega_l^S-\omega_k^D-\omega)+(1-f_l^S)f_k^D\delta(\omega_l^S-\omega_k^D+\omega)),
\ee 
which takes the expression given in the main text in the continuous limit of the reservoirs' spectra, we finally obtain:
\bb
{\cal L}_\text{QPC}\rho
&=&  - i[H'_{LS},\rho]+\sum_{\omega=0,\pm\Omega}\gamma_\text{QPC}(\omega)D[\hat n_\omega]\rho \,, 
\ee
where 
\bb 
H'_{LS} &=& \sum_{\omega=0,\pm\Omega} \sum_{kl}\chi_{kl}^2\left(f_l^S(1-f_k^D)F_{\Delta t}(\omega_l^S-\omega_k^D-\omega)-(1-f_l^S)f_k^D F_{\Delta t}(\omega_l^S-\omega_k^D+\omega)\right)\hat n_\omega^\dagger \hat n_\omega \nonumber\\
&&+2 \sum_{kl}\chi_{kl}g_{kl}\left(f_l^S(1-f_k^D)-(1-f_l^S)f_k^D\right)F_{\Delta t}(\omega_l^S-\omega_k^D)\hat n_0
\ee
is a small correction to the double-dot energy transitions due to the coupling to the QPC, which is omitted in the main text as it can be included as a re-normalization of $\Omega$.

\subsection{Energy exchanges within the QPC}\label{s:App_QPC_energy}

We assume that the source has a chemical potential $\mu_M > 0$ while the drain Fermi energy is used as the energy reference. We can define power $P_M^\text{in}$ injected to fuel the QPC, and the heat flows exchanged with the source $J_S$ and drain $J_D$ as:
\bb
P_M &=& -\mu_M  \frac{d}{dt}\moy{  \sum_q s_q^\dagger s_q}\\
J_S &=& -  \frac{d}{dt}\moy{  \sum_q (\omega^S_q - \mu_M) s_q^\dagger s_q}\\
J_D &=& -  \frac{d}{dt}\moy{  \sum_p \omega^D_p  d_p^\dagger d_p }
\ee

\subsubsection{Work flow}

\bb
P_M &=& -\mu_M \text{Tr}_{S,D,qd}\left\{ \sum_q s_q^\dagger s_q  \frac{\Delta \rho_\text{tot}^I}{\Delta t}\right\}\nonumber\\
&=& \mu_M \sum_q  \frac{1}{\Delta t}\int_{t}^{t+\Delta t} dt' \int_{t}^{t'} dt'' \text{Tr}\Big\{ s_q^\dagger s_q  [\Upsilon(t') + \Xi(t') \hat n_R(t'),[\Upsilon(t'') + \Xi(t'') \hat n_R(t''),\rho_\text{tot}^I(t'')]\Big\}\nonumber\\
&=& \mu_M \sum_q  \frac{1}{\Delta t}\int_{t}^{t+\Delta t} dt' \int_{0}^{t'-\tau} d\tau \Big[ \moy{[s_q ^\dagger s_q,\Upsilon(t') + \hat n_R(t') \Xi(t')](\Upsilon(t'-\tau) + \hat n_R (t'-\tau)\Xi(t'-\tau))}  + \text{H.c.}\Big)\Big]\;\;\;
\ee

We now expand $\hat n_R(t) = \sum_\omega \hat n_\omega^\dagger e^{-i\omega t}$ and apply as before the secular approximation, keeping terms which do not oscillate as a function of $t'$. Indeed, those terms all oscillate at a frequency much larger than $\Delta t^{-1} \ll \Omega$. We define 
\bb
j_{qk}(\omega) &=& 
\frac{1}{\Delta t}\text{Re}\int_{t}^{t+\Delta t} dt' \int_{0}^{t'-\tau}d\tau \left[f^S_q(1-f^D_k) e^{i(\omega^S_q-\omega^D_k)\tau}- (1-f^S_q)f^D_k e^{-i(\omega^S_q-\omega^D_k)\tau}\right] e^{-i \omega \tau}\nonumber\\ &=& G_{\Delta t}(\omega_q^S-\omega_k^D-\omega)f^S_q(1-f^D_k)-G_{\Delta t}(\omega_k^D-\omega_q^S-\omega)(1-f^S_q)f^D_k
\ee
to express the non-zero contributions:
\bb
\frac{1}{\Delta t}\text{Re}\int_{t}^{t+\Delta t} dt' \int_{0}^{t'-\tau} d\tau\text{Tr}\Big\{ [s_q^\dagger s_q,\Upsilon]\Upsilon(-\tau)\rho(t)\otimes\rho^\text{eq}_{S,D}\Big\} &=& \sum_{k}g_{qk}^2  j_{qk}(0)\nonumber\\
\frac{1}{\Delta t}\text{Re}\int_{t}^{t+\Delta t} dt' \int_{0}^{t'-\tau} d\tau\text{Tr}\Big\{ [s_q^\dagger s_q, \Xi]\Xi  (-\tau) \hat n_\omega^\dagger  \hat n_\omega e^{-i\omega\tau}\rho(t)\otimes\rho^\text{eq}_{S,D}\Big\}  &=&  \sum_{k}\chi_{qk}^2 j_{qk}(\omega)\moy{\hat n_\omega^\dagger \hat  n_\omega}\nonumber\\
\frac{1}{\Delta t}\text{Re}\int_{t}^{t+\Delta t} dt' \int_{0}^{t'-\tau} d\tau\text{Tr}\Big\{ [s_q^\dagger s_q, \Xi]\Upsilon(-\tau) \hat  n_0 \rho(t)\otimes\rho^\text{eq}_{S,D}\Big\}  &=&  \sum_{k}\chi_{qk}g_{qk}  j_{qk}(0) \moy{\hat n_0}\nonumber\\
\frac{1}{\Delta t}\text{Re}\int_{t}^{t+\Delta t} dt' \int_{0}^{t'-\tau} d\tau\text{Tr}\Big\{ [s_q^\dagger s_q, \Upsilon]\Xi(-\tau) \hat n_0 \rho(t)\otimes\rho^\text{eq}_{S,D}\Big\}  &=&  \sum_{k}\chi_{qk}g_{qk}  j_{qk}(0)  \moy{\hat n_0}
\ee

We obtain:
\bb
P_M
&\simeq& 2\mu_M\sum_{qk}\Big[\moy{(g_{qk} + \chi_{qk} \hat n_0)^2}  j_{qk}(0) + \chi_{qk}^2\sum_{\omega=\pm\Omega} \moy{\hat n_\omega^\dagger\hat  n_\omega}  j_{qk}(\omega)\Big]
\ee
As before, we use that $G_{\Delta t}(u) = \text{Re}A_{\Delta t}(u)$ is much narrower than $f_k^{S,D}(u)$ in the considered regime, which allows us to take the continuous limit of the source and drain spectra:
\bb\label{eqS:PandI}
P_M&\simeq& \mu_M I_\text{QPC},\quad\quad I_\text{QPC}= \sum_{\omega=0,\pm \Omega} I_\omega
\ee
where:
\bb\label{eqS:QPCcurrentcontrib}
I_0&=& \int dE {\cal F}_0(E) \left\langle\left(\sqrt{{\cal T}_0(E)} - \sqrt{{\cal T}_M(E)}\hat n_0\right)\left(\sqrt{{\cal T}_0(E)} - \sqrt{{\cal T}_M(E)}\hat n_0\right)\right\rangle,\nonumber\\
I_{\pm\Omega}&=&  \int dE {\cal T}_M(E){\cal F}_\omega(E)  \moy{\hat n_{\pm\Omega}^\dagger \hat n_{\pm\Omega}},
\ee
and 
\bb
{\cal F}_\omega(E) &=& f_S(E)(1-f_D(E-\omega))-(1-f_S(E))f_D(E+\omega),
\ee
and the QPC transparencies (see Eq.\eqref{d:T01M}). We have assumed that ${\cal T}_1 \leq {\cal T}_0$, such that $\sqrt{{\cal T}_M} = \sqrt{{\cal T}_0}-\sqrt{{\cal T}_1}$ \cite{Goan01}.

\subsubsection{Dissipated heat flow}

With the same approach and approximations, we find:

\bb
J_S &=& - \text{Tr}_{SD,qd}\left\{ \sum_q (\omega_q^S-\mu_M) s_q^\dagger s_q   \frac{\Delta \rho_\text{tot}^I}{\Delta t}\right\}\nonumber\\
&=&  \sum_q  (\omega_q^S-\mu_M) \frac{1}{\Delta t}\int_{t}^{t+\Delta t} dt' \int_{t}^{t'} dt'' \text{Tr}\Big\{ s_q^\dagger s_q  [\Upsilon(t') + \Xi(t') \hat n_R(t'),[\Upsilon(t'') + \Xi(t'') \hat n_R(t''),\rho_\text{tot}^I(t'')]\Big\}\nonumber\\
&=& 2\sum_{qk}  (\omega_q^S-\mu_M) \Big[\moy{(g_{qk} + \chi_{qk} n_0)^2}  j_{qk}(0) + \chi_{qk}^2\sum_{\omega=\pm\Omega} \moy{n_\omega^\dagger n_\omega}  j_{qk}(\omega)\Big],
\ee
and
\bb
J_D &=& - \text{Tr}_{SD,qd}\left\{ \sum_p \omega_p^D d_p^\dagger d_p   \frac{\Delta \rho_\text{tot}^I}{\Delta t}\right\}\nonumber\\
&=&  \sum_p  \omega_p^D \frac{1}{\Delta t}\int_{t}^{t+\Delta t} dt' \int_{t}^{t'} dt'' \text{Tr}\Big\{ d_p^\dagger d_p  [\Upsilon(t') + \Xi(t') \hat n_R(t'),[\Upsilon(t'') + \Xi(t'') \hat n_R(t''),\rho_\text{tot}^I(t'')]\Big\}\nonumber\\
&=& -2\sum_{qk} \omega_k^D  \Big[\moy{(g_{qk} + \chi_{qk} n_0)^2}  j_{qk}(0) + \chi_{qk}^2\sum_{\omega=\pm\Omega} \moy{n_\omega^\dagger n_\omega}  j_{qk}(\omega)\Big]
\ee

The total heat flow dissipated by the measuring apparatus then corresponds to 
\bb
J_M = J_S+J_D\,.
\ee

\subsubsection{Measurement-induced energy flow}

The total energy provided by the QPC reservoirs is therefore:
\bb
J_M + P_M &=&
2\sum_{qk} (\omega_q^S-\omega_k^D)  \Big[\moy{(g_{qk} + \chi_{qk} n_0)^2}  j_{qk}(0) + \chi_{qk}^2\sum_{\omega=\pm\Omega} \moy{n_\omega^\dagger n_\omega}  j_{qk}(\omega)\Big]\ee
As before, we use that $G_{\Delta t}(u) = \text{Re}A_{\Delta t}(u)$ is much narrower than $f_k^{S,D}(u)$ to replace it with a Dirac delta function. As a consequence, the factor $(\omega_k^S-\omega_q^D)j_{qk}(0)$ vanishes:
\bb
\nonumber\\
 J_M + P_M  &=& 2\sum_{qk}\chi_{qk}^2\sum_{\omega=0,\pm\Omega} \omega \moy{n_\omega^\dagger n_\omega}  \left[f_q^S(1-f_k^D)\delta(\omega_q^S-\omega_k^D-\omega)+(1-f_q^S)f_k^D\delta(\omega_k^S-\omega_q^D-\omega)\right]\nonumber\\
 &=& \sum_{\omega=0,\pm\Omega} \omega \gamma_\text{QPC}(\omega) \moy{n_\omega^\dagger n_\omega} \nonumber\\
  &=& Tr[H_\text{QD}{\cal L}_M\rho_\text{qd}] \nonumber\\
  &=& \dot{E}_M.
\ee

\subsection{QPC shot noise and signal to noise ratio}\label{s:App_QPC_SNR}

\subsubsection{QPC current and activity}

As a starting point to evaluate the electric current noise in the QPC, we decompose the evolution of the total system during $\Delta t$ in terms of jump events between the source and drain reservoirs (which contribute to the QPC current and shot noise), and evolution between such jumps events \cite{Blasi2024,Landi24}. Because of the coupling to the double-dot system, each electron transfer is accompanied by an effect on the density operator $\rho$ which is captured by a superoperator. More precisely, we define:
\bb
{\cal L}^{+}[\rho] &=&  \frac{1}{\Delta t}\text{Tr}_{S,D}\left\{\int_{t}^{t+\Delta t} dt' \int_{t}^{t'} dt''V_+^I(t')\rho_\text{tot}^I(t'')V_+^{I\dagger}(t'')\right\}+\text{H.c.}.\\
{\cal L}^{-}[\rho] &=&  \frac{1}{\Delta t}\text{Tr}_{S,D}\left\{\int_{t}^{t+\Delta t} dt' \int_{t}^{t'} dt''V_-^I(t')\rho_\text{tot}^I(t'')V_-^{I\dagger}(t'')\right\}+\text{H.c.}\nonumber\\
{\cal L}^\text{nj}[\rho] &=& \frac{\Delta\rho^I_\text{tot}}{\Delta t} - {\cal L}^{+}[\rho_\text{tot}]-{\cal L}^{-}[\rho_\text{tot}].
\ee
where 
\bb
V_+^I(t) = \sum_{qp} \left(g_{qp}+\hat{n}_R(t)\chi_{qp}\right) s_q^\dagger d_p e^{i(\omega^S_q-\omega^D_p)t}\\
V_-^I(t) = \sum_{qp} \left(g_{qp}+\hat{n}_R(t)\chi_{qp}\right)  d_p^\dagger s_q e^{-i(\omega^S_q-\omega^D_p)t},
\ee
are the terms of the coupling operator $V^I(t)$ associated with an electron transfer from the source to the drain (+) and from the drain to the source (-), which verifies $V_+^I(t)+V_-^I(t)=V^I(t)$.\\

Using the Fourier decomposition of $\hat{n}_R(t)$, one can further decompose the superoperators ${\cal L}^\pm$ into components associated with a transition of energy $\hbar\omega \in \{0,\pm\hbar\Omega\}$ in the double-dot system:
\bb
{\cal L}^\pm &=& \sum_{\omega=0,\pm\Omega} {\cal L}^\pm_\omega\quad\quad\text{with}\quad\quad{\cal L}^\pm_\omega[\rho] = \int dE \;{\cal F}_\omega^\pm(E)\;\; \hat K_\omega(E)\rho \hat K_\omega^{\dagger}(E).
\ee
They are defined in terms of the Kraus operators:
\bb
\hat K_0 &=& \sqrt{{\cal T}_0(E)}-\sqrt{{\cal T}_M(E)}\hat n_0,\quad\quad
\hat K_\Omega = \sqrt{{\cal T}_M(E)}\hat n_\Omega,\quad\quad
\hat K_{-\Omega} =\sqrt{{\cal T}_M(E)}\hat n_{-\Omega},
\ee
and the functions
\bb
{\cal F}^+_\omega(E) &=& f_S(E)(1-f_D(E-\omega)),\quad\quad {\cal F}^-_\omega(E)=(1-f_S(E))f_D(E+\omega)
\ee

We can now express the superoperators measuring the current  ${\cal I}$, and the activity ${\cal A}$ in the QPC:
\bb 
{\cal I}= {\cal L}^+-{\cal L}^-\quad\quad\text{and}\quad\quad {\cal A}= {\cal L}^++{\cal L}^-.
\ee
They are related to the average current and activity, given that the dots are in the state $\rho$ via $I_\text{QPC}(\rho) = \text{Tr}\{{\cal I}[\rho]\}$, $A_\text{QPC}(\rho) = \text{Tr}\{{\cal A}[\rho]\}$.

\subsubsection{Quantum regression approach}

The zero-frequency current noise at steady state is defined as:
\bb 
s_{II}(\nu = 0) = \int_{-\infty}^{\infty} d\tau S_{II}(\tau),
\ee
where $S_{II}(\tau)$ is the steady-state auto-correlation function of the QPC current. The latter is related to the current and activity operators via \cite{Blasi2024}:
\bb
S_{II}(\tau) = \delta(\tau)A_\text{ss} + \underbrace{\text{Tr}\left\{{\cal I}e^{{\cal L}_\text{tot}|\tau|}{\cal I}[\rho_\text{ss}]\right\} - I_{QPC}^2}_{C_{II}(\tau)}.
\ee
To compute the term $C_{II}(\tau)$, we resort to the Quantum Regression formula \cite{CarmichaelBook}.

We first note that the projectors on the energy eigenstates of the dot system constitute a complete set of observables obeying the set of differential equations:
\bb
\forall i\in\{+,-\},\quad\quad\moy{\dot{\Pi}_i(t)} = \sum_{j=\pm} A_{ij} \moy{\Pi_j(t)}+B_i,
\ee
where $\Pi_\pm = \ket{\pm}\bra{\pm}$. The matrix ${\bf A}$ and the vector are determined by the master equation \eqref{eq:ME}, using the closure relation $\ket{00}\bra{00}+\Pi_++\Pi_-=\idop$ to eliminate the population of state $\ket{00}$.
The Quantum Regression Theorem then states that the steady-state correlation functions $C_{\cal O}^{(j)}(\tau)=\text{Tr}\{\Pi_j e^{{\cal L}_\text{qd}\tau}{\cal O}[\rho_\text{ss}]\}$, for any superoperator ${\cal O}$, obey the coupled set of differential equations:
\bb
\forall i\in\{+,-\},\quad\quad\frac{d}{d\tau} C_{\cal O}^{(i)}(\tau) = \sum_{j=0,\pm} A_{ij} C_{\cal O}^{(j)}(\tau) + \moy{\cal O}_\text{ss} B_i,
\ee
with $\moy{{\cal O}}_\text{ss}=\text{Tr}\{{\cal O}[\rho_\text{ss}]\}$.
Introducing the column vector $\vec C_{\cal O}(\tau) = \left(C_{\cal O}^{(+)}(\tau),C_{\cal O}^{(-)}(\tau)\right)^T$, one then has:
\bb 
\vec C_{\cal O}(\tau) = \moy{\cal O}_\text{ss}\left(e^{A\tau}\left(\frac{\vec C_{\cal O}(0)}{\moy{\cal O}_\text{ss}}-\vec P_\text{ss}\right)+\vec P_\text{ss}\right),
\ee
where $\vec P_\text{ss}=-{\bf A}^{-1}\vec B = \left(p_{+,ss},p_{-,ss}\right)^T$ is the vector of steady-state populations.

Moreover, we note that for any operator $X = \sum_i x_i \Pi_i$, we have $\text{Tr}\{{\cal I}X\} = \sum_{k=0,\pm} i_k \text{Tr}\{\Pi_k X\}$, with:
\bb 
i_0 &=& \int dE \left({\cal F}^+_0(E)-{\cal F}^-_0(E)\right){\cal T}_0(E),\\
i_+ &=& \int dE \left[\left({\cal F}^+_0(E)-{\cal F}^-_0(E)\right)\left(\sqrt{{\cal T}_0(E)}c^2+\sqrt{{\cal T}_1(E)}s^2\right)^2+\left({\cal F}^+_{-\Omega}(E)-{\cal F}^-_{-\Omega}(E)\right){\cal T}_M(E)c^2s^2\right]\\
i_- &=& \int dE \left[\left({\cal F}^+_0(E)-{\cal F}^-_0(E)\right)\left(\sqrt{{\cal T}_0(E)}s^2+\sqrt{{\cal T}_1(E)}c^2\right)^2+\left({\cal F}^+_{\Omega}(E)-{\cal F}_{\Omega}^-(E)\right){\cal T}_M(E)c^2s^2\right].
\ee
Consequently, we can express the QPC current auto-correlation function in terms of $\vec C_{\cal I}(\tau)$ (using once more the closure relation):
\bb
C_{II}(\tau) =  I_{QPC}\left(i_0 + \vec i \cdot \frac{\vec C_{\cal I}(\tau)}{\moy{I}_\text{ss}}\right) - I_{QPC}^2= I_{QPC}\left(i_0 + \vec i \cdot\left[e^{{\bf A}\tau}\left(\frac{\vec C_{\cal I}(0)}{{\moy{I}_\text{ss}}}-\vec P_\text{ss}\right)+\vec P_\text{ss}\right]  - I_{QPC}\right).
\ee
where $\vec i = (i_0-i_+,i_0-i_-)$. We finally use that $i_0+\vec i\cdot\vec P_\text{ss} = I_{QPC}$ to deduce the expression of the QPC shot noise:
\bb
s_{II}(\nu = 0) &=& A_\text{QPC}-I_\text{QPC} \vec i \cdot {\bf A}^{-1}\left(\frac{\vec C_{\cal I}(0)}{{\moy{I}_\text{ss}}}-\vec P_\text{ss}\right)\nonumber\\
 &=&\left(a_0+\vec a\cdot \vec P_\text{ss}\right)  -\vec i \cdot {\bf A}^{-1} \left(\begin{array}{c}
     \left((i_+-i_0)-\moy{I}_\text{ss}\right)p_{+,ss} \\ \left((i_--i_0)-\moy{I}_\text{ss}\right)p_{-,ss}.
 \end{array}\right)
 \ee
 In the last line, we have introduced the vector $\vec a = (a_+-a_0,a_--a_0)^T$ and the scalars $a_0$, $a_\pm$ built on the same model as $\vec i, i_0$ and $i_\pm$ to evaluate the activity:
\bb 
a_0 &=& \int dE \left({\cal F}^+_0(E)+{\cal F}^-_0(E)\right){\cal T}_0(E),\\
a_+ &=& \int dE \left[\left({\cal F}^+_0(E)+{\cal F}^-_0(E)\right)\left({\cal T}_0(E)c^2+{\cal T}_1(E)s^2\right)^2+\left({\cal F}^+_{-\Omega}(E)+{\cal F}^-_{-\Omega}(E)\right){\cal T}_M(E)c^2s^2\right]\\
a_- &=& \int dE \left[\left({\cal F}^+_0(E)+{\cal F}^-_0(E)\right)\left({\cal T}_0(E)s^2+{\cal T}_1(E)c^2\right)^2+\left({\cal F}^+_{\Omega}(E)+{\cal F}_{\Omega}^-(E)\right){\cal T}_M(E)c^2s^2\right].
\ee

\subsubsection{Signal-to-noise ratio}

We evaluate the signal-to-noise ratio by dividing the difference between the two values of the signal $\Delta I_\text{QPC}$ (the electric current through the QPC) and the associated noise $S_{II}(0)$.
Precisely:
\bb 
\Delta I_\text{QPC} &=& I_\text{QPC}^{0}-I_\text{QPC}^{1}\nonumber
\ee
where $I_\text{QPC}^{0}$ and $I_\text{QPC}^{1}$ are the values of the current associated with the right dot being empty or occupied, respectively. Those values are obtained from Eqs.~\eqref{eqS:PandI}-\eqref{eqS:QPCcurrentcontrib}, by noting that when the right dot is occupied, we have:
\bb 
\bra{+}\rho\ket{+} = \cos^2\theta,\quad\quad \bra{-}\rho\ket{-} = \sin^2\theta,\quad\quad (\moy{\hat n_R}=0)\quad\quad\quad\quad\\
\moy{\hat n_0} = 2\cos^2\theta\sin^2\theta,\quad\quad \moy{\hat n^\dagger_\Omega \hat n_\Omega} = \cos^4\theta \sin^2\theta, \quad\quad \moy{\hat n^\dagger_{-\Omega} \hat n_{-\Omega}} = \cos^2\theta \sin^4\theta
\ee
In contrast, when the right dot is empty, the state of the double dot system is not completely determined and can be parametrized by the population $p_{00}$ of the state $\ket{00}$:
\bb 
\bra{+}\rho\ket{+} = \sin^2\theta(1-p_{00}),\quad\quad \bra{-}\rho\ket{-} = \cos^2\theta(1-p_{00}),\quad\quad (\moy{\hat n_R}=1)\quad\quad\quad\quad\\
\moy{\hat n_0} = 1-p_{00},\quad\quad \moy{\hat n^\dagger_\Omega \hat n_\Omega} = \cos^2\theta \sin^4\theta(1-p_{00}), \quad\quad \moy{\hat n^\dagger_{-\Omega} \hat n_{-\Omega}} = \cos^4\theta \sin2\theta(1-p_{00}).
\ee
To evaluate the worst-case signal-to-noise ratio, we select the value of $p_{00}$ corresponding to the smallest signal variation $\Delta I_\text{QPC}$, which turns out to be $p_{00}=0$.

%\bibliography{biblio}

\end{document}